\documentclass{article}
\usepackage{graphicx} 
\usepackage{amsmath} 

\title{The Duality of Whittaker Potential Theory: Fundamental Representations of Electromagnetism and Gravity, and Their Orthogonality}
\author{Mark Titleman}
\date{October 2026}

\begin{document}

\maketitle
E. T. Whittaker produced two papers in 1903 and 1904 that, although sometimes considered mere mathematical statements (Barrett, 1993), held important implications for physical theory. The Whittaker 1903 paper united electrostatic and gravitational attraction as resulting from longitudinal waves – waves whose wavefronts propagate parallel to their direction. The Whittaker 1904 paper showed that electromagnetic waves resulted from the interference of two such longitudinal waves or scalar potential functions. Although unexplored, the implications of these papers are profound: gravitational lensing, gravitational waves, the Aharonov-Bohm effect, the existence of a hyperspace above or behind normal space, the elimination of gravitational and point charge singularities, MOND, and the expansion of the universe. This last implication can be related to the recent finding that black holes with posited vacuum energy interior solutions alongside cosmological boundaries have a cosmological coupling constant of k=3, meaning that black holes gain mass proportionally to $a^{3}$ in a parameterization equation within a Robertson-Walker cosmology and are a cosmological accelerated expansion species (Farrah et al., 2023). This expansion and many features of General Relativity can be explained by the mass-proportionality and preferred direction of the longitudinal waves within the two underlying non-local Whittaker potentials (Titleman, 2022). Expansion of the universe is produced as longitudinal motion within the Whittaker potentials only when dynamic electromagnetism is separate from time-static gravity in intergalactic space.

\section{Introduction}
Modern theories of gravity have faced several unresolved difficulties such as unexplained expansion, failure to adhere to predicted galactic rotations curves, the existence of unphysical gravitational singularities, and incompatibility with Quantum theory. It is thus useful to assess older classical theories of gravity that overtly or implicitly offered several features of Relativity. E. T. Whittaker's 1903 paper on partial differential equations anticipated General Relativity in many ways by proposing an undulatory theory of gravity and a static gravitational field resulting from propagating effects – the field is the result of electromagnetic processes. Whittaker’s 1904 paper on two scalar potential functions showed that the electromagnetic four-potential of Relativity overlooked other forms of electromagnetic potential. Even though no action could be set up for computing local physical processes, Whittaker potential theory foresaw the Aharonov-Bohm effect and could be used to replace Dirac spinors in the Dirac equation (Ruse, 1937). What the cautious Whittaker considered an “undulatory theory” could in fact explain several features of Relativity in addition to MOND. For example, gravitational lensing can be understood as resulting from the preferred direction of the non-local potentials and their mass-proportionality (Titleman, 2022). Finally, due to the dynamic longitudinal motion in the z-axis being additive, Whittaker potential theory can also provide a simple explanation for expansion of the universe - it is merely dynamic light decoupled from static gravity and can only be produced intergalactically.

\section{Whittaker Potential Theory}
E. T. Whittaker's 1903 paper on partial differential equations found a harmonic solution to the amenable central equations of calculus in three dimensions: the wave equation and the more specific Laplace equation. Both potentials could be analysed into simple plane waves, bringing new unity to potential theory, inviting the possibility of new physical phenomena, and implying the modern notion that calculus alone is insufficient as a physical theory. The question of whether potentials – the second derivative of which produces force fields – are real is irrelevant to such possibilities and their confirmation via other observational data and new mathematical techniques. Whittaker considered “gravitation and electrostatic attraction explained as modes of wave-disturbance” (Whittaker, 1903), meaning that the force fields associated with matter and charge are undulatory and perhaps matter and charge themselves. Whittaker’s 1903 paper thus displayed incredible predictive power in recognizing the wave nature of force and force carrier (matter, charge). Its mathematical generality and novelty were reported in the British popular press - which was an impressive feat for a young mathematician - yet it was far too ahead of its time for an uneventful period in British physics with little observational data and only a recent introduction of continental mathematics.
\\
\\
According to Whittaker, an electrostatic or gravitational field, varying with the inverse square of distance, results from waves propagating at any speed and in any which way. The general solution to the Laplace equation was found in the form:
\begin{equation} \label{eq1}
\int_{0}^{2\pi}f(xcosv+ysinv+iz, v)dv
\end{equation}
where f is an arbitrary function of the two arguments. This is accomplished in terms of Bessel functions by expanding the function f as a Taylor series with respect to the first argument and a Fourier series with respect to the second argument. The v is a periodic argument. Using similar analysis, the general solution to the wave equation was found as dynamic and longitudinal in the form:
\begin{equation} \label{eq2}
\int_{0}^{\pi}\int_{0}^{2\pi}f(xsinucosv+ysinusinv+zcosu+ct,u,v)dudv
\end{equation}
where f is an arbitrary function of the three arguments.
\\
\\
Regarding statics, Whittaker’s 1903 paper claimed that once longitudinal waves interfere with each other the disturbance at any point does not depend on time but only on position. Force potential can therefore be defined in terms of standing waves (non-local solution) as well as propagating waves (local solution changing in time) (Barrett, 1993). This undulatory theory of gravity propagating with a finite velocity subsumed gravity to the transmission of electromagnetic radiation, forming a significant contribution to the electromagnetic worldview of the day. The aether producing longitudinal as well as transverse electromagnetic waves was a common belief of 19th century physicists and was given a mathematically detailed treatment by Whittaker (Carvalo \& Rodrigues, 2008).
\\
\\
Hector Munera (Munera, 2018) writes that Whittaker's claim of generality is unacceptable due to periodicity being assumed, although this permitted Whittaker's analysis. Munera also emphasizes that equation (1) implies a rotation in the complex plane $(z,ixcos\theta+iysin\theta)$. A time-dependent solution in the form of equation (2) is realized by “projecting z and $xcos\theta+ysin\theta$ onto ray r directed at angle $\phi$ relative to the Z-axis, thus shifting to spherical coordinates” (Munera, 2018).
\\
\\
Whittaker's 1904 paper on two scalar potentials showed that electromagnetic fields could be decomposed into two scalar potential functions as intersecting beams, with orthogonal sphericity when discontinuous (gravitational). Whittaker accomplished this by defining three scalar fields and eliminating the orthogonal sphericity term using a gauge. The two scalar potentials F and G derive the magnetic force h and dielectric displacement d as:
\\
\[d_{x}=\frac{\partial^2F}{\partial{x}\partial{z}}+\frac{1}{c}\frac{\partial^2G}{\partial{y}\partial{t}}\]

\[d_{y}=\frac{\partial^2F}{\partial{y}\partial{z}}-\frac{1}{c}\frac{\partial^2G}{\partial{x}\partial{t}}\]

\[d_{z}=\frac{\partial^2F}{\partial{z^2}}-\frac{1}{c^2}\frac{\partial^2G}{\partial{t^2}}\]

\[h_{x}=\frac{1}{c}\frac{\partial^2F}{\partial{y}\partial{t}}-\frac{\partial^2G}{\partial{x}\partial{z}}\]

\[h_{y}=-\frac{1}{c}\frac{\partial^2F}{\partial{x}\partial{t}}-\frac{\partial^2G}{\partial{y}\partial{z}}\]

\[h_{z}=\frac{\partial^2G}{\partial{x^2}}+\frac{\partial^2G}{\partial{y^2}}\tag{3}\]

F and G are represented asymmetrically as follows:
\\
\[F(x,y,z,t)=\sum\frac{e}{4\pi}sinh^{-1}\frac{\bar{z}'-z}{((\bar{x}'-x)^2+(\bar{y}'-y)^2)^{1/2}}\]
\[G(x,y,z,t)=\sum\frac{e}{4\pi}tan^{-1}\frac{\bar{y}'-y}{\bar{x}'-x}\tag{4}\]
\\
The summation is taken over the electrons in the field. In continuous form they are:
\\
\[F=\int_{0}^{\pi}\int_{0}^{2\pi}f(xsinucosv+ysinusinv+zcosu+ct,u,v)dudv\]
\[G=\int_{0}^{\pi}\int_{0}^{2\pi}g(xsinucosv+ysinusinv+zcosu+ct,u,v)dudv\tag{5}\]
\\
The shift between spherical (polar) and planar (Cartesian) coordinates can be seen in Whittaker’s 1951 representation of two scalar potentials F and G:
\\
\[F(x,y,z,t)=\frac{1}{2}\sum_{e}log\frac{\bar{r}+\bar{z}'-z}{\bar{r}-(\bar{z}'-z)}\]
\[G(x,y,z,t)=-i\frac{1}{2}\sum_{e}log\frac{\bar{x}'-x+i(\bar{y}'-y)}{\bar{x}'-x-i(\bar{y}'-y)}\tag{6}\]
\\
Whittaker wrote:
\begin{quote}
   It will be noted that F and G are defined in terms of the positions of the electrons alone, and do not explicitly involve their velocities. Since in the above formulae for d and h an interchange of electric and magnetic quantities corresponds to a change of G into F and of F into G, it is clear that the two functions F and G exhibit the duality which is characteristic of electromagnetic theory: thus an electrostatic field can be described by F alone, and a magnetostatic field by G alone; again, if the field consists of a plane wave of light, then the functions F and G correspond respectively to two plane-polarised components into which it can be resolved. Since there are an infinite number of ways of resolving a plane wave of light into two plane-polarised components, it is natural to expect that, corresponding to any given electromagnetic field, there should be an infinite number of pairs of functions F and G capable of describing it, their difference from each other depending on the choice of the axes of co-ordinates-as is in fact the case. Thus there is a physical reason why any particular pair of functions F and G should be specially related to one co-ordinate, and cannot be described by formulae symmetrically related to the three co-ordinates (x,y,z) (McCrea, 1952).
\end{quote}

\section{A New Explanation for Expansion of the Universe}
This new understanding of waves at the interface of plane and spherical rotations strongly suggests the mathematical and physical concept of vorticity as dynamic 3-parameter space. It is clear from the Whittaker analysis that a more massive observer would experience more longitudinal waves than only the two experienced by an observer as an electromagnetic wave, and it is implicit that when observed at the speed of light the number of longitudinal waves would collapse into the orthogonal axis (non-local). Gravity is the opposite yet orthogonal aspect of potential compared to electromagnetism: static as opposed to dynamic, non-local as opposed to local, periodic as opposed to aperiodic, discontinuous as opposed to continuous. Whittaker's analysis, for example the 1951 representation, clearly opened mathematical possibilities beyond calculus and natural logarithms. Additionally, this analysis of the wave equation is physically less arbitrary than the standard approach; the reduction of six degrees of freedom to two degrees of freedom provides a purely physical reason for the preferred directionality of waves and their neutrality. Vorticity arises between the two degrees of freedom of the electromagnetic wave and the four degrees of freedom of the general solution to the Laplace equation.
\\
\\
The free parameters assigned to the axes of a wave within a vorticity are as follows: longitudinal motion or speed in the z is charge-proportional from the perspective of the observer (compressible potentials), number of longitudinal waves is mass-proportional from the perspective of the observer and folds into the non-local (static) when observed at high speed (local vorticity force), and the x-axis or plane wave axis is related to amplitude, intensity, and soliton radius.
\\
\\
Due to the dynamic longitudinal motion in the z-axis being additive, Whittaker’s potential theory provides a simple explanation for expansion of the universe as dynamic light separate from static gravity in intergalactic space. If this is the case, there would be an inverse relation – with implicit coordinate shift – between the coupled amplitude or changing background intensity of the universe and expansion of the universe. Operations such as antiderivative, tetration (which can be given more mathematical centrality), or antiderivative followed by tetration can be performed on this relation. Since the intensity of the universe is double that of all predicted stars (Lauer et al., 2022), the relation would be on the order of 3/2.
\\
\[\sqrt\frac{L_{v}}{2} = \frac {Expansion\ (yz-plane)}3 \tag{7}\]
\\
\[\int3\sqrt{\frac{L_{v}}{2}}dL_v = {Expansion\ (yz-plane)}\]
\[\sqrt2(\frac{L_{v}}{2})^{3/2} = {Interpolation}\]
\[^{\infty}[\sqrt2](\frac{L_{v}}{2})^{3/2} = L_{v}\sqrt\frac{L_{v}}{2} = \Psi_\bot\tag{8}\]
\hfill (Wadell, 1935)
\\
\\
This can be considered a form of computation. $\sqrt{2}$ is subtended and angular in the context of infinite tetration, orthogonal sphericity, and associated 3+1 manifolds. The question “how many squares fit in a sphere?" is analogous to the relevant question in Ramsey theory “how big must some structure be to guarantee that a particular property holds?" 
\\
\\
The “3” is the result of the new interpretation of three dimensions or three axes permitted by this interpretation of Whittaker potential theory. Longitudinal waves are additive in two directions – phase and antiphase z-directions (F,G). Black holes produce these longitudinal waves as scalar potentials, providing cosmological coupling, a third additive “direction” (non-local), another dynamic component, and an important center of wave decomposition for scalar potentials and vorticity ($\Psi$). This understanding replaces black hole singularities with vacuum energy interior solutions (or densities) within a Robertson-Walker cosmology, as elaborated on by Whittaker (Whittaker, 1935) and Farrah (Farrah et al., 2023).

\section{Three Methods for Computing the MOND Acceleration Constant}
\textbf{Quantum theory (Boltzmann-Rydberg statistics)}
\\
If black holes produce longitudinal waves as scalar potentials, it is via beam splitting within scalar interferometry. This reduces the four degrees of freedom inherent to Whittaker’s general solution of the Laplace equation (x, y, z, non-local Z) to the two local degrees of freedom inherent to Whittaker’s general solution of the wave equation. 
\\
\\
This halving can also be understood in statistical terms as normality. Black holes keep absolute time as a simple geometry while statistics and ultimately probability are primordial. Critical density is traditionally arrived at by adjusting the Hubble parameter and - indeed - the scale factor of the universe is the inverse mathematical and physical operation of equation (7). 
\\
\[a(t) = (\frac {t_ \frac{1}{2}}{t})^\frac {2}{3}\tag{9}\]
\\
Can the average kinetic energy of the cosmic microwave background be measurable in terms of half time $k_b$$\Delta$T and related to a non-local "force" applied on black hole-containing galaxies - the “Whittaker force potential force” (vorticity force) - towards the computation of the MOND acceleration constant? This is due to the four degrees of freedom of this latent kinetic energy becoming two upon splitting within a black hole. The following equation was previously proposed by the author (Titleman, 2020):
\\
\[\frac{\alpha R_\infty k_b \Delta T}{m_p}=1.21*10^{-10} m/s^2\tag{10}\]
\\
Although dynamics and statics and the generality of Whittaker potentials replace mass-energy and calculus, a second layer of mass as the Planck mass (homogeneous, theoretical) – as well as a second layer of brightness-dependence as the fine structure constant scaled to galactic brightness and shape – may provide the MOND acceleration constant and the Tully-Fischer relation.
\\
\\
\textbf{Ternary probability}
\\
The gauge used by Whittaker to eliminate the orthogonal sphericity term ($\Psi$) and reduce the electromagnetic wave from six degrees of freedom to two degrees of freedom (F,G) is equivalent with the following statement under the assumptions of calculus and mass-energy:
\\
\[\sqrt3\ x = \ x^{-\sqrt3}\]
\[x = 0.818\]
\\
\[G - 0.818G\]
\[= 1.214 * 10^{-11} \frac{m^{3}}{{s^{2}{kg}}}\tag{11}\]
\\
Eliminating $\Psi$ is equivalent with gaining an order of magnitude and adding $G$, producing the MOND acceleration constant and permitting coordinate shifting and ternary probability.
\\
\\
\\
\textbf{The cosmological constant and Cartesian asymmetry}
\\
The relation of the MOND acceleration constant to the cosmological constant has been ascertained as:
\\
\[a_0=\sqrt\frac{\Lambda}{3}\tag{12}\]
\\
According to this new understanding of three axes, the mass-proportional, static gravitational non-local axis is related to the charge-proportional, dynamic electromagnetic z-axis by squaring. Two directions of dynamism are in the z and a third occurs in the non-local axis as black hole growth (occurs in all directions locally). Static gravity is only in the mass-proportional and thus limited-range observed non-local axis. The potentials are non-local in most senses. As such, the dynamism at the interface of the cosmologically coupled z-axis and observed (as a wave) non-local axis are related by squaring only within the limited range of nearby matter. Outside of this limited range there is simply expansion of the universe. Squaring must also be used for the cosmological constant in the context of spacetime – where the interface between dynamic z-axis and static non-local axis is constantly implied. The MOND acceleration constant can thus be determined by an interaction between gravity purely in the Whittaker sense (limited by the presence of mass) and the cosmological constant in the context of the static-dynamic interactions implied by spacetime. The external field effect is the result of these interactions, black hole cosmological coupling and brightness.

\section{Implications of Whittaker's General Solutions and Two-Scalar Potential Theory}
The Cartesian asymmetry and shift between spherical and planar coordinates of Whittaker's analysis give rise to many physical implications. The z-direction (propagation direction) is necessarily treated differently than the other two spatial directions. Gravity thus manifests solely in the purely orthogonal or non-local direction due to the preferred direction of the potentials and their mass-proportionality. Gravitational lensing results from the two scalar potentials interfering with each other with mass as a free parameter.
\\
\\
These papers described gravity and electromagnetism not only as modes of disturbance in the same medium, but as providing a broad mathematical explanation for gravitational waves. Whittaker claimed that this potential theory provided for an “undulatory theory of gravity” (Whittaker, 1903).
\\
\\
Whittaker potential theory anticipated the Aharonov-Bohm effect since the potentials F and G are considered more basic entities. Fields require the potentials to exist, but potentials can exist on their own and produce phenomena such as the Aharonov-Bohm effect – a particle affected electromagnetically without electric or magnetic fields present.
\\
\\
Finally, virtually all singularities can be eliminated by this theory. A point charge as one type of singularity would not need to exist. Charge appears collectively as longitudinal motion carrying radiation. The need for a propagation medium for transverse waves was in fact predicted by Maxwell and other classical physicists since they consist of orthogonal electric and magnetic waves, the former being undulating dipolar electric fields that were considered to require separated and opposite electric charges. Massless charge as the free parameter of motion in the two scalar potentials partially inverts this belief, but accords with Maxwell’s dielectric medium while simplifying Maxwell's findings from a physical perspective and doing away with non-existent point charges. Gravitational singularities vanish as well. Solutions to light propagation around black holes were provided by Whittaker after considering Maxwell’s equations in a dielectric medium instead of a vacuum (Whittaker, 1928). Black holes can be viewed similarly to the two scalar potentials - although complex and three-dimensional - and may work collectively as charge does, exist as part of the hyperspatial structure as orthogonal sphericity, and generate the Whittaker potentials (localized around each galaxy’s supermassive black hole) through wave decomposition.

\section{Conclusion}
This understanding of the analytical papers of E. T. Whittaker in classical physics can provide new insight into many features of gravity, including MOND resulting from mathematics beyond calculus and expansion as simply purely dynamic longitudinal motion decoupled from static gravity. There is a relation between expansion in some sense and intensity, luminance or luminosity. Finally, two-scalar potential may explain the relation between the cosmological constant in the context of spacetime and the MOND acceleration constant, or the MOND acceleration constant generally.
\\
\\
Whittaker had introduced continental math, including a rigorous treatment of the Laplace equation and associated equations and a novel implementation of Bessel functions, in the early 20th century when British mathematics had become stagnant. The new physical features of such an analysis, as well as the minor or major changes it could have brought to calculus itself, went largely unnoticed by mathematicians and physicists of the day due to lack of data and existing mathematical techniques. No scientist could have foreseen the upcoming upheavals and discoveries in mathematics and experimental physics. Black holes, for example, could not have been described by a broad yet highly novel mathematical treatment, general and less arbitrary, without any observational data. 
\\
\\
Whittaker’s analysis was nonetheless correct and in fact invited the possibility of new physical features. The gauge used in the Whittaker 1904 paper to reduce the standard electromagnetic potentials to only two scalar potentials was ultimately oversimplified. It can be expanded through advances in computation and the Wick rotation which already links statistical mechanics to quantum mechanics and 4D Euclidean space to spacetime. Ternary probability, interpolation, ray tracing, statistical mechanics and information are of central importance. A language of Clifford algebra or the geometry of a Clifford torus (with luminosity and a black hole network phase space, for incorporating General Relativity) can be developed. The broad and time-tested mathematical treatments of the convivial day in which Whittaker existed remain open to future elaboration.

\section{References}
Barrett, T. W. (1993). Electromagnetic phenomena not explained by Maxwell's equations. \textit{In Essays on the formal aspects of electromagnetic theory.} (pp. 6-86).
\\
\\
Farrah, D., Croker, K. S., Zevin, M., Tarlé, G., Faraoni, V., Petty, S., Weiner, J. (2023). Observational evidence for cosmological coupling of black holes and its implications. \textit{The Astrophysical Journal Letters}, 944(2), L31.
\\
\\
Lauer, T. R., Postman, M., Spencer, J. R., Weaver, H. A., Stern, S. A., Gladstone, G. R., Young, L. A. (2022). Anomalous flux in the cosmic optical background detected with new horizons observations. \textit{The Astrophysical Journal Letters}, 927(1), L8.
\\
\\
McCrea, W. H. (1952). \textit{History of Theories of the Aether and Electricity.} I. By Sir Edmund Whittaker Pp. xiv, 434. 32s. 6d. 1951.(Nelson). The Mathematical Gazette, 36(316), 138-141.
\\
\\
Múnera, H. A. (2018). Neo-Cartesian unified fluid theory: from the classical wave equation to De Broglie’s Lorentzian quantized mechanics and quantized gravity. \textit{In Unified Field Mechanics II: Formulations and Empirical Tests: Proceedings of the Xth Symposium Honoring Noted French Mathematical Physicsist Jean-Pierre Vigier}, Porto Novo, Italy, 25-28 July 2016 (pp. 198-220).
\\
\\
Ruse, H. S. (1937). On Whittaker’s Electromagnetic ‘Scalar Potentials’. \textit{The Quarterly Journal of Mathematics}, (1), 148-160.
\\
\\
Titleman, M. (2020). Gravitation Due to Scalar Potentials and Black Holes. \textit{Physics International}, 11(1), 1-3.
\\
\\
Titleman, M. (2022). Representations and Implications of Papers Written by ET Whittaker in 1903 and 1904. \textit{arXiv preprint} arXiv:2205.08309.
\\
\\
Trovon de Carvalho, A. L., Rodrigues Jr, W. A. (2001). \textit{The non sequitur mathematics and physics of the “new electrodynamics” proposed by the AIAS group.}
\\
\\
Wadell, H. "Volume, shape, and roundness of quartz particles." \textit {The Journal of geology} 43, no. 3 (1935): 250-280.
\\
\\
Whittaker, E. T. (1904). On an expression of the electromagnetic field due to electrons by means of two scalar potential functions. \textit{Proc. Lond. Math. Soc, 1}, 367.
\\
\\
Whittaker, E. T. (1903). On the partial differential equations of mathematical physics. \textit{Mathematische Annalen}, 57(3), 333-355.
\\
\\
Whittaker, E. T. (1935). On Gauss' theorem and the concept of mass in general relativity. \textit{Proceedings of the Royal Society of London. Series A, Mathematical and Physical Sciences}, 149(867), 384-395.
\\
\\
Whittaker, E. T. (1928). The influence of gravitation on electromagnetic phenomena. \textit{Journal of the London Mathematical Society}, 1(2), 137-144.
\end{document}